\documentclass[a4paper,10pt]{article}

\usepackage[utf8]{inputenc}
\usepackage{amsmath}
\usepackage{amsfonts}
\usepackage{amssymb}
\usepackage{fontenc}
\usepackage{mathrsfs}

\usepackage{parskip}

\renewcommand{\Im}{\,{\rm Im}\,}
\renewcommand{\Re}{\,{\rm Re}\,}

\newcommand{\upd}{{\rm d}}

 \def\Co{{\mathbb C}}
 
 \def\Io{{\mathbb I}}
 \def\Mo{{\mathbb M}}
 \def\Ro{{\mathbb R}}
 
 \def\hh{{\bf h}}

 \def\kk{{\bf k}}
 \def\Ee{{\bf E}}
 \def\Bb{{\bf B}}
 \def\ranglec{\rangle^{\mbox{\tiny c}}}
 \def\langlec{\strut^{\mbox{\tiny c}}\hskip -1pt\langle}
\newcommand{\url}[1]{\tt#1}

\newtheorem{proposition}{Proposition}

\title{Gauge transformations of a relativistic field of quantum harmonic oscillators}

\author{
Jan Naudts\\
        Departement Fysica, Universiteit Antwerpen,\\
        Universiteitsplein 1, 2610 Antwerpen, Belgium\\
        e-mail: \url{Jan.Naudts@uantwerpen.be}\\
        orcid: \url{https://orcid.org/0000-0002-4646-1190}
}

\begin{document}

\maketitle

\begin{abstract}
A set of gauge transformations of a relativistic field of quantum harmonic oscillators
is studied in a mathematically rigorous manner.
Each wave function in the domain of the number operator of a single oscillator
generates a Fr\'echet-differentiable
field of wave functions.
Starting from a coherent wave function one obtains a two-dimensional differentiable
manifold of coherent vector states.
As an illustration it is shown
that the gauge transformation can be chosen in such a way that the
resulting fields describe a freely-propagating wave.
\end{abstract}


\section{Introduction}

The goal of the present paper is to give a rigorous treatment of a
set of gauge transformations in the case of a simple model of quantum field theory.

Gauge transformations are useful to derive quantum field theories from first principles.
Originally, they were considered as a means to eliminate unwanted degrees of freedom
from Maxwell's theory of Electrodynamics.
Only later \cite{GD92} it became clear
that they form an important group of symmetries,
together with the group of Poincar\'e transformations.
With the birth of Yang-Mills theories \cite {YM54} gauge invariance arguments
started being used to formulate quantum field theories.
The discovery of the Berry phase \cite {BMV84}
introduced gauge transformations as unitary
operators in a well-defined Hilbert space of wave functions.
Incentive for the present paper is a transformation from the lab frame to the moving
frame, in the way it is done in a recent review \cite{KSMP17} of applications of gauge theory
in Solid State Physics. 

The Hilbert space under consideration is that of the quantum harmonic oscillator.
In the above mentioned review \cite{KSMP17} this model is worked out as an example.
Here, it is treated in a relativistic setting.
The gauge transformations are composed of unitary operators $U(x)$ 
that depend on space time position $x$.
They are generated by the four elements of the Lie
algebra of the oscillator group \cite{SRF67}:
the creation and annihilation operators $a^\dagger$ and $a$,
the number operator $a^\dagger a$ and the identity $\Io$.
The gauge potential $A_\mu(x)$ can be identified with the electromagnetic vector potential.
It is proved in the present work that this vector potential is
a uniquely-defined self-adjoint operator. 

The present work is part of an effort to reformulate Quantum Electrodynamics (QED)
in a rigorous manner by abandoning the requirement that the representation of the
canonical commutation relations should be irreducible. Czachor and coworkers studied
reducible QED in a series of papers. 
See \cite{CM00,CM04,CN06,CW09} and references given there.
The present author introduced a simplified formalism \cite{NJ17,NJ19b,NJ19a}.
By starting from gauge transformations, as is done here,
the possibility of a rigorously formulated QED can be
investigated in a more systematic manner.

\subsection*{Notations}

Greek indices $\mu,\nu,\rho,\sigma$ have values 0,1,2,3 and follow the Einstein summation convention.
They refer to components of the space time position.
On the other hand, the index $\alpha$ refers to spatial components
and take on the values 1,2,3, without summation convention.
Throughout the text raising and lowering indices involves the Minkowski pseudo metric
$g$ with sign convention $(+,-,-,-)$.
The  following abbreviation is used
\begin{equation}
\partial_\mu\equiv\frac{\partial\,\,\,}{\partial x^\mu}.
\nonumber\end{equation}

Capitals such as $A$, $B$, $\cdots$ are operators in the Hilbert space
of the quantum harmonic oscillator.
A notational exception are the creation and annihilation operators $a^\dagger$ and $a$.
Vectors in $\Ro^3$ are set in boldface.

Wave functions, elements of $\mathscr H$, are denoted $\psi$, $\zeta$. 
The inner product $(\psi,\zeta)$ of $\psi$ and $\zeta$ is also written as $\langle\zeta|\psi\rangle$.
It is linear in $\psi$ and anti-linear in $\zeta$.

\subsection*{Structure of the paper}

Quantities such as wave functions in Hilbert space or operators on Hilbert space,
when depending on a space time position $x$, are quantum fields. 
This point of view is shortly discussed in the next section. 
Section \ref {sect:disp} reviews the definition of the displacement operator
and some of its properties. Section \ref {sect:diffield} introduces
transformations from a lab frame to a moving frame in the spirit of
\cite{KSMP17}. With such a transformation is associated a vector of gauge potentials $R_\mu(x)$
that are self-adjoint operators on $\mathscr H$.
Section \ref {sect:geom} considers a field of states on the algebra of 
bounded operators on the Hilbert space $\mathscr H$. The set of states reachable
from a given state by a Lorentz transform and corresponding gauge transformation
forms a differentiable manifold. Tangent planes are discussed in Section \ref {sect:tangent}.
Section \ref {sect:field} introduces the generalized
anti-symmetric field tensor.
Section \ref{sect:wavevector} illustrates the present approach by characterizing
fields with electric and magnetic vectors orthogonal to a given wave vector $\kk\not=0$.
The final section contains a short discussion of the results.
Technical proofs are gathered in Appendix.

\section{Fields of Hilbert spaces}

Let $\Mo$ denote Minkowski space. Let ${\mathscr H}$ denote the Hilbert space of the
quantum harmonic oscillator.
Following Dixmier \cite{DJ69,DJ81}
a continuous map $x\in \Mo\mapsto \zeta(x)\in{\mathscr H}$ is a
{\em vector field} over $\Mo$, element of a {\em continuous field of Hilbert spaces}.
Nowadays it is called a Hilbert space bundle over $\Mo$.

The space of vector fields is denoted ${\mathscr G}$.
The subspace of bounded continuous maps is denoted ${\mathscr B}$.
It is a Banach space for the norm
\begin{equation}
||\zeta||=\sup_x||\zeta(x)||.
\nonumber\end{equation}
A special subspace is formed by the constant maps. Any $\zeta$ in ${\mathscr H}$ can be identified
with the constant map $x\mapsto\zeta$ in ${\mathscr B}$. 
Hence, one has ${\mathscr H}\subset{\mathscr B}\subset {\mathscr G}$.

Any bounded linear operator $T$ on ${\mathscr H}$ defines a continuous linear
operator, denoted $\hat T$ and working on ${\mathscr B}$, by 
\begin{equation}
(\hat T\zeta)(x)=T\zeta(x),
\qquad x\in \Mo, \zeta\in {\mathscr B}.
\nonumber\end{equation}

A {\em diagonal} operator $\hat T$ on ${\mathscr B}$ is defined by a map $x\in\Mo\mapsto T(x)$,
where $T(x)$ is a linear operator on ${\mathscr H}$, together with a domain $\mathscr D$,
subspace of ${\mathscr B}$, such that for each $\zeta$ in $\mathscr D$
the map $x\in\Mo\mapsto T(x)\zeta(x)$ is continuous.

If $T$ is a closed operator on ${\mathscr H}$ then $\hat T$ is a closed operator on ${\mathscr B}$
with domain ${\mathscr D}$ consisting of the elements $x\mapsto\zeta(x)$ of ${\mathscr B}$
with all $\zeta(x)$ in the domain of $T$.

Of special interest for the present paper are diagonal
operators $x\in\Mo\mapsto U(x)$ where each operator $U(x)$
is a unitary transformation of ${\mathscr H}$. Any strongly continuous map $x\in\Mo\mapsto U(x)$
into the isometries of ${\mathscr H}$ defines an isometry $\hat U$ of ${\mathscr B}$.

For proofs see Appendix 1 of \cite{NJ19a}.

\section{The displacement operator}
\label{sect:disp}

The displacement operator $D(z)$, with $z$ a complex number, is the unitary operator
defined by \cite {GJP09}
\begin{equation}
D(z)=\exp(za^\dagger-\overline z a).
\nonumber\end{equation}
Here, $a$ is the annihilation operator working in the Hilbert space of a single
quantum harmonic oscillator. For a rigorous definition see Appendix \ref{app:Weyl}.

Let $|0\rangle$ denote the ground state of the harmonic oscillator.
The coherent state  $|z\ranglec$
is obtained from the vacuum state $|0\ranglec=|0\rangle$ by the action of the displacement
operator
\begin{equation}
|z\ranglec=D(z)|0\rangle.
\nonumber\end{equation}
The composition of two displacement operators satisfies the relation
\begin{equation}
D(z_1)D(z_2)=e^{-i(z_1\wedge z_2)}D(z_1+z_2)
\nonumber\end{equation}
with 
\begin{equation}
z_1\wedge z_2=z_1'z_2''-z_1''z_2'=\frac i2\left(z_1\overline z_2-\overline z_1 z_2\right).
\nonumber\end{equation}
Here, $z'$ and $z''$ denote the real, respectively imaginary part of $z$.

Common knowledge about coherent wave functions is that 
\begin{equation}
a|z\ranglec=z|z\ranglec
\quad\mbox{ and }
\exp(ir a^\dagger a)|z\ranglec=|e^{ir}z\ranglec
\end{equation}
for any real number $r$.
Note that $|z\ranglec$ belongs to the domain of the number operator $a^\dagger a$.
This follows from
\begin{equation}
||a^\dagger a |z\ranglec||^2=e^{-|z|^2}\sum_{n=0}^\infty \frac{1}{n!}n^2|z|^{2n}
=|z|^2(1+|z|^2)<+\infty.
\nonumber\end{equation}

For further use note the following.

\begin{proposition}{Proposition} \label{disp:prop:deriv}
Assume $f$ is a differentiable complex-valued function.
For any $\zeta$ in the domain of the number operator $a^\dagger a$
the Fr\'echet derivative of $x\mapsto D(f(x))\zeta$ exists and is given by
\begin{equation}
i\hbar \partial_\mu D(f(x))\zeta
=X_\mu(x)D(f(x))\zeta
\label{calc:result2}
\end{equation}
with $X_\mu(x)$ given by
\begin{equation}
\frac 1\hbar
X_\mu(x)=
f(x)\wedge \partial_\mu f(x)
+i\partial_\mu f(x)\,a^\dagger
-i\partial_\mu \overline{f(x)}\,a.
\label{disp:Xdef}
\end{equation}
\end{proposition}

The proof is found in Appendix \ref {app:proofprop}.

\section{A differentiable field}
\label{sect:diffield}

Fix a complex-valued function $f(x)$ and real-valued functions $\phi(x)$ and $\gamma(x)$
with $x$ in $\Mo$. Assume these functions are continuous with continuous derivatives.
Consider unitary operators $U(x)$ defined by
\begin{equation}
U(x)=e^{i\phi(x)}D(f(x))^\dagger e^{i\gamma(x)a^\dagger a}D(f(x)).
\label{diff:Udef2}
\end{equation}

One can use $D(z)^\dagger a D(z)=a+z$ to write
\begin{equation}
U(x)=e^{i\phi(x)}e^{i\gamma(x)(a+f(x))^\dagger(a+f(x))}.
\label{diff:Udef3}
\end{equation}
Note that the set of unitary operators of this form is not closed under multiplication.
The set of infinitesimal generators forms a Lie algebra \cite {SRF67}.
It contains the Heisenberg algebra as a subalgebra.

Choose also a normalized wave function $\zeta$ in $\mathscr H$.
A wave function $\eta(x)$ is defined by
\begin{equation}
\eta(x)=U(x)\zeta.
\nonumber\end{equation}
Following \cite{KSMP17}, $\eta(x)$ is interpreted as a wave function in a {\em moving frame}
while $\zeta$ is a wave function in the Hilbert space at space time position $x=0$.

For simplicity of notation introduce $\eta$ given by $\eta=\eta(0)=U(0)\zeta$.

\begin{proposition}{Proposition} \label{diff:prop:coh}
The wave function $\eta(x)=U(x)\zeta$ is coherent up to a complex phase factor if and only if
$\zeta$ is coherent up to a complex phase factor.
\end{proposition}

This is shown by explicit calculation. See the Appendix \ref {app:coherent}.

It follows from Proposition \ref {disp:prop:deriv} that the field
$x\mapsto \eta(x)$ is Fr\'echet-differentiable.
Indeed, one can prove the following.

\begin{proposition}{Proposition} \label{gauge:prop:difffield}
Let ${\mathscr D}$ denote the domain of the number operator $a^\dagger a$.
One has
\begin{description}
 \item [\quad (a) \quad]
If $\zeta$ is in ${\mathscr D}$
then the field $x\mapsto\eta(x)$ is Fr\'echet-differentiable.
 \item [\quad (b) \quad]
 If $\partial_\mu \gamma(x)\not=0$ then
a self-adjoint operator $R_\mu(x)$ with domain $\mathscr D$ is defined by
\begin{eqnarray}
R_\mu(x)&=&
-\hbar\,\partial_\mu \phi(x) \cr
&-&
\hbar\, (\partial_\mu \gamma(x))\, (a+f(x))^\dagger(a+f(x))\cr
&+&
i\hbar\left(e^{i\gamma(x)}-1\right)(\partial _\mu f(x))\,(a+f(x))^\dagger\cr
&-&
i\hbar\left(e^{-i\gamma(x)}-1\right)(\partial_\mu \overline {f(x)})\,(a+f(x)).
\label{diff:Rpot}
\end{eqnarray}
It satisfies
\begin{equation}
i\hbar\, \partial_\mu \eta(x)
=R_\mu(x)\eta(x),
\qquad\zeta\in{\mathscr D}.
\nonumber\end{equation}
\end{description}
\end{proposition}
The proof is given in Appendix \ref{app:proofprop2}.

Consider a Lorentz transformation
\begin{equation}
\tilde x^\mu=\Lambda^\mu_{\,\,\,\nu}x^\nu.
\nonumber\end{equation}
The vector potential $R_\mu$, as given by (\ref{diff:Rpot})
transforms as a covector
\begin{equation}
R_\nu(x)=\Lambda^\mu_{\,\,\,\nu}\tilde R_\mu(\tilde x),
\nonumber\end{equation}
where the vector $\tilde R_\mu$ is defined starting from $\tilde U(\tilde x)=U(x)$
and is obtained from (\ref{diff:Rpot})
by transforming the functions $\phi$, $\gamma$, and $f$ accordingly
($\tilde\phi(\tilde x)=\phi(x)$ and similar for $\gamma$ and $f$).

\section{A manifold of vector states}
\label{sect:geom}

Fix a normalized wave function $\zeta$ in the domain of the number operator $a^\dagger a$.
The field of wave functions $x\mapsto\zeta(x)$ is defined by
\begin{equation}
\zeta(x)=U^\dagger (x)U(0)\zeta=U^\dagger(x)\eta.
\nonumber\end{equation}
It belongs to the Banach space $\mathscr B$.
Note that $\zeta(0)=\zeta$. 
The reasoning behind this definition is that, given a wave vector $\zeta$
at the lab position $x=0$ the transformation $U(0)$ is needed to obtain the wave vector $\eta$,
which belongs to the moving frame. Next a transformation back yields
the value $\zeta(x)$ in a frame with space time position $x$.

There exist self-adjoint operators $A_\mu(x)$ such that
\begin{equation}
i\hbar\partial_\mu\zeta(x)=-A_\mu\zeta(x), \zeta\in{\mathscr D}.
\nonumber\end{equation}
They are related to the operators $R_\mu(x)$ by
\begin{equation}
A_\mu(x)=U^\dagger(x)R_\mu(x)U(x).
\end{equation}
From (\ref {diff:Udef3}) it is clear that $U^\dagger(x)$ is obtained from $U(x)$
by inverting the sign of the functions $\phi$ and $\gamma$.
Hence, an explicit expression for $A_\mu(x)$ obtained from (\ref{diff:Rpot}) reads
\begin{eqnarray}
A_\mu(x)&=&
-\hbar\,\partial_\mu \phi(x) \cr
& &
-\hbar\, (\partial_\mu \gamma(x))\, (a+f(x))^\dagger(a+f(x))\cr
& &
-i\hbar\left(e^{-i\gamma(x)}-1\right)(\partial _\mu f(x))\,(a+f(x))^\dagger\cr
& &
+i\hbar\left(e^{i\gamma(x)}-1\right)(\partial_\mu \overline {f(x)})\,(a+f(x)).
\label{diff:Apot}
\end{eqnarray}


The quantum expectation of a bounded operator $Y$ on the Hilbert space $\mathscr H$
given the field $\zeta$ is denoted $x\mapsto \omega_{x}(Y)$ and is given by
\begin{equation}
\omega_{x}(Y)=(Y\zeta(x),\zeta(x))=\langle \zeta(x)|Y\zeta(x)\rangle.
\nonumber\end{equation}
For each $x$ is $\omega_{x}$ a vector state on the von Neumann algebra
of bounded operators on $\mathscr H$.
Hence, $x\mapsto \omega_{x}$ is a field of vector states.
It defines a manifold $M_\zeta$ by
\begin{equation}
M_\zeta=\{\omega_x:\,x\in\Mo\}.
\nonumber\end{equation}

Let unitary operators $V_{y,x}$ be defined by 
\begin{equation}
V_{y,x}=U^\dagger(y)U(x),
\qquad x,y\in\Mo.
\nonumber\end{equation}
One has $V_{z,y}V_{y,x}=V_{z,x}$ and $V_{x,y}V_{y,x}=V_{x,x}=\Io$.
Hence, $(x,y)\mapsto V_{y,x}$ is a unitary representation of the groupoid of directed arcs in
Minkowski space $\Mo$.

The gauge transformation $V_{y,x}$, when acting on the field of wave functions $x\mapsto \zeta(x)$, gives
\begin{equation}
V_{y,x}\zeta(x)=\zeta(y),
\qquad x,y\in\Mo.
\nonumber\end{equation}
This induces a translation groupoid of actions $\tau_{y,x}$ on the manifold $M_\zeta$
defined by
\begin{equation}
\tau_{y,x}\omega_x=\omega_y,
\qquad x,y\in\Mo.
\nonumber\end{equation}

One expects the manifold $M_\zeta$ to be three-dimensional because it depends on a real-valued
function $\gamma(x)$ and a complex-valued function $f(x)$.
However, if $\zeta$ is a coherent wave function then by Proposition \ref {diff:prop:coh}
there exists a coherent wave function $|z(x)\ranglec$ such that
$\omega_x(Y)=\langlec z(x)|Y|z(x)\ranglec$ for all $Y$.
Hence, the manifold depends on a single complex function $z(x)$ and is at most two-dimensional.

It is easy to see that the map $\omega_x\mapsto z(x)$ is well-defined.
Hence, it is a chart for the manifold $M_\zeta$. Indeed, if $\omega_x=\omega_y$
then it follows that $\zeta(x)=e^{i\xi}\zeta(y)$ for some real number $\xi$.
This implies that $|z(x)\ranglec$ is a multiple of $|z(y)\ranglec$. The latter 
is only possible if $z(x)=z(y)$.

\section{Tangent plane}
\label{sect:tangent}

The  vectors tangent to the field $x\mapsto\omega_{x}$ at space time position $x$
are linear functionals given by
\begin{equation}
\chi_{y,x}(Y)
=
\frac{\upd\,\,\,}{\upd\lambda}\omega_{x+\lambda y}(Y)\bigg|_{\lambda=0}.
\nonumber\end{equation}
One has
\begin{eqnarray}
\frac{\upd\,\,\,}{\upd\lambda}\omega_{x+\lambda y}(Y)
&=&
\left(Y\zeta(x+\lambda y),\frac{\upd\,\,\,}{\upd\lambda}\zeta(x+\lambda y)\right)\cr
& &
+\left(Y\frac{\upd\,\,\,}{\upd\lambda}\zeta(x+\lambda y),\zeta(x+\lambda y)\right)\cr
&=&\frac i\hbar y^\mu
\left([Y,A_\mu(x+\lambda y)]_{_-} \zeta(x+\lambda y),\zeta(x+\lambda y)\right).
\nonumber\end{eqnarray}
Let the action of an operator $Z$
on a functional $\chi$ be defined by
\begin{equation}
\check Z\,\chi(Y)=i\chi([Y,Z]_{_-}).
\nonumber\end{equation}
The above expression then becomes
\begin{equation}
\frac{\upd\,\,\,}{\upd\lambda}\omega_{x+\lambda y}
=
\frac 1\hbar y^\mu \check A_\mu(x+\lambda y)
\omega_{x+\lambda y}.
\label{geo:eqmot}
\end{equation}
%
The tangent vector $\chi_{y,x}$ is given by
\begin{equation}
\chi_{y,x}
=
\frac 1\hbar y^\mu \check A_\mu(x)\omega_{x}.
\nonumber\end{equation}
The covariant derivatives $\check D_\mu$ are given by           
\begin{equation}
\check D_\mu
=
\mathring \partial_\mu+\frac 1\hbar \check A_\mu(x)
\nonumber\end{equation}
with $\mathring \partial_\mu$ defined by $\mathring \partial_\mu\chi(Y)=\chi(\partial_\mu Y)$.

From expression (\ref {diff:Apot}) it is clear that the tangent vector $\chi_{y,x}$
is a real-linear combination of three functionals
\begin{eqnarray}                                    
e^{(1)}_x(Y)&=& \check a^\dagger \check a\,\omega_x(Y)\cr
e^{(2)}_x(Y)&=& [\check a^\dagger +\check a]\,\omega_{x}(Y)\cr
e^{(3)}_x(Y)&=&-i[\check a^\dagger -\check a]\,\omega_{x}(Y).
\nonumber\end{eqnarray}
Hence the tangent space is at most three-dimensional.
A short calculation shows that if $\zeta(x)$ is a coherent wave function then one has
\begin{equation}
e^{(1)}_x=(\Re z(x))e^{(2)}_x-(\Im z(x))e^{(3)}_x   
\nonumber\end{equation}
with $\zeta(x)$ proportional to $|z(x)\ranglec$.
In particular, if $\zeta(x)$ is the wave function of the ground state
of the harmonic oscillator then one has $e^{(1)}_x=0$.

\section{Field equations}
\label{sect:field}

The generalized anti-symmetric field tensor $\check F$ is defined by
\begin{equation}
\check F_{\mu\nu}= \hbar \left[\check D_\mu,\check D_\nu\right]_{_-}. 
\nonumber\end{equation}
When acting on a linear functional $\chi$ one obtains 
\begin{equation}
\check F_{\mu\nu}\chi=\chi\left(i[Y,F_{\mu,\nu}]_{_-}\right)
\nonumber\end{equation}
with
\begin{equation}                                                       
F_{\mu\nu}(x)
=
\partial_\mu  A_\nu(x)-\partial_\nu  A_\mu(x)
-\frac{i}{\hbar}\left[ A_\mu(x), A_\nu(x)\right]_{_-}
\nonumber\end{equation}
A short calculation which uses $U(x)A_\mu(x)=i\hbar\partial_\mu U(x)$ gives 
\begin{eqnarray}
\partial_\mu  A_\nu(x)
&=&
\partial_\mu\, \left(
U^\dagger(x){i\hbar}\partial_\nu\, U(x)\right)\cr
&=&
{i\hbar} \left(\partial_\mu\,
U^\dagger(x)\right)\partial_\nu\, U(x)
+
{i\hbar}U^\dagger(x)\partial_\mu\partial_\nu U(x)\cr
&=&
\frac{i}{\hbar}A_\mu(x)A_\nu(x)+i\hbar U^\dagger(x)\partial_\mu\partial_\nu U(x).
\nonumber\end{eqnarray}
This shows that $F_{\mu\nu}(x)=0$. 
In particular, the generalized force tensor $\check F_{\mu\nu}(x)$ vanishes.
This shows up in the above calculations as a cancellation of the 
traditional contribution $\partial_\mu  A_\nu(x)-\partial_\nu  A_\mu(x)$
with the contribution due to the non-commutatitvity of the components $A_\mu(x)$
of the vector potential.

The canceling contributions to $F_{\mu\nu}(x)$ are denoted $\mathring F_{\mu\nu}(x)$
\begin{equation}
\mathring F_{\mu\nu}(x)
=
\partial_\mu \,A_\nu-\partial_\nu \,A_\mu=\frac{i}{\hbar}\left[ A_\mu(x), A_\nu(x)\right]_{_-}.
\nonumber\end{equation}
From (\ref {diff:Apot}) one obtains             
\begin{eqnarray}
\mathring F_{\mu\nu}(x)
&=&
-i\hbar\left(e^{i\gamma(x)}-1\right)\left(e^{-i\gamma(x)}-1\right)\,\left[
\partial_\mu f(x)\,\partial_\nu \overline{f(x)}\,
-
\partial_\nu f(x)\,\partial_\mu \overline{f(x)}\,
\right]\cr
& &
-\hbar\left(e^{-i\gamma(x)}-1\right)\,
\left[
\partial_\mu\gamma(x)\,\partial_\nu f(x)\,
-
\partial_\nu\gamma(x)\,\partial_\mu f(x)\,
\right]\,
(a+f(x))^\dagger
\cr
& &
-\hbar\left(e^{i\gamma(x)}-1\right)\,
\left[
\partial_\mu\gamma(x)\,\partial_\nu \overline{f(x)}\,
-
\partial_\nu\gamma(x)\,\partial_\mu \overline{f(x)}\,
\right]\,
(a+f(x)).\cr
& &
\label {diff:antisym}
\end{eqnarray}

It is straightforward to verify that the anti-symmetric tensor $\mathring F_{\mu\nu}$ as
given by (\ref {diff:antisym}) satisfies the Bianchi identities
\begin{equation}
\partial_\sigma\,\mathring F_{\mu,\nu}(x)
+
\partial_\mu\,\mathring F_{\nu,\sigma}(x)
+
\partial_\nu\,\mathring F_{\sigma,\mu}(x)
=0.
\nonumber\end{equation}
In particular, the fields $\mathring F_{\mu\nu}$ can be interpreted as the components
of electric and magnetic field vectors satisfying Maxwell's equations.

\section{Wave vector dependence}
\label{sect:wavevector}

The 3-dimensional vector  $\Ee$ with components
$E_\alpha=\mathring F_{0\alpha}$, $\alpha=1,2,3,$ can be interpreted as an
electric field vector. Similarly the components $(\mathring F_{32},\mathring F_{13},\mathring F_{21})$
are the components $(B_1,B_2,B_3)$ of a magnetic field vector $\Bb$.
The present section investigates whether the functions 
$\phi$, $\gamma$ and $f$ introduced in Section \ref {sect:diffield}
can be chosen in such a way that these fields are orthogonal
to a given vector $\kk\not=0$, called the {\em wave vector}.

Use $\nabla f$ to denote the 3-dimensional vector with components $\partial_\alpha f(x)$,
and similar for $\nabla \gamma$.
As usual, the convention $k_0=|\kk|$ is used so that $k_\mu k^\mu=0$.
The following result is proved in Appendix \ref{app:ortho}.

\begin{proposition}{Proposition} 
Let be given a 3-dimensional wave vector $\kk\not=0$.
Non-trivial solutions of
\begin{equation}
  \Ee\cdot\kk=\Bb\cdot\kk=0
  \label{wave:orthocond}
\end{equation}
can exist if the functions $f$ and $\gamma$ are related by
\begin{equation}
\nabla f
=
z\nabla\gamma+w\kk.
\nonumber\end{equation}
and
\begin{equation}
\partial_0 f(x)
=
\left(z+\frac{w|\kk|^2}{\kk\cdot\nabla\gamma}\right)
\partial_0 \gamma(x),
\label{app:ortho:timecond}
\end{equation}
where, $z$ and $w$ are complex constants.

\end{proposition}

Explicit expressions for $\Ee$ and $\Bb$ are found in the Appendix.
The electric field $\Ee$ is parallel to $|\kk|^2\nabla\gamma-(\kk\cdot\nabla\gamma)\kk$,
the magnetic field $\Bb$ is parallel to $\kk\times\nabla\gamma$.

\section{Discussion}

The paper focuses on the Hilbert space $\mathscr H$ of the quantum harmonic oscillator. 
Transformations from a lab frame to a moving frame and back result in a quantum field
over space time position $x$. The composed transformation is by definition the
gauge transformation. It accompanies the Lorentz transformation from space time position
$x=0$ to space time position $x$. From this point of view the gauge transformation
implements parallel transport between different lab frames and
normalized wave functions describe the states of the quantum system
in a tangent plane located at the position of the observer.

The generators of the gauge transformations are requested
to belong to the Lie algebra of the oscillator group.
With this restriction calculations remain tractable. 

It is shown in Proposition \ref {gauge:prop:difffield} that 
if the wave function $\zeta$ belongs to the domain of the number operator then the
transformation from lab frame to moving frame is Fr\'echet-differentiable.
The same proposition shows that the gauge potential operators $A_\mu(x)$
are uniquely-defined self-adjoint operators. 

The manifold $M_\zeta$ of vector states consisting of the wave functions $\zeta(x)$,
obtained from a single wave function $\zeta$ by gauge transformations from $x=0$ to 
space time position $x$,
and its tangent planes are considered in Sections \ref{sect:geom}, 
respectively \ref{sect:tangent}.
If the wave function $\zeta$ 
is coherent then $M_\zeta$ is a two-dimensional differentiable manifold of coherent states.

As an illustration it is shown in Section \ref{sect:wavevector} that
the formalism can describe fields with electric and magnetic vectors orthogonal to
a given wave vector $\kk\not=0$.

Let us emphasize some aspects of the present approach.

The transformations from lab frame to moving frame do not form a group.
The composition of a transformation to the moving frame followed by a transformation
back to a lab frame at a different space time position is 
a gauge transformation. 
These gauge transformations form a unitary representation of the groupoid of
directed arrows in Minkowski space.
A single wave function $\zeta$ in the Hilbert space of the
observer is transformed to another frame of observation by a gauge transformation
that accompanies the Lorentz transformation. This leads to the interpretation of
a quantum field as a way to consider a single quantum state in all possible frames of
observation.

In the conventional approach the gauge potential arises from the application of
the {\em gauge principle} to global symmetries of a matter field. In the present approach
no matter field is around. 
The bosonic states of the harmonic oscillator generate the force field. 
The gauge transformation is parameterized by a real function $\gamma(x)$ and a complex function $f(x)$.
This gives a large freedom to adapt the model to specific needs.
It is not clear for the moment how to make use of this freedom.

The specific choice of unitary transformations is justified by the argument that they
map coherent wave functions into coherent wave functions. The restriction to
coherent fields reduces the dimension of the manifold from 3 to 2 and facilitates
detailed calculations.

\appendix
\section*{Appendix}

\section{Weyl operators}
\label{app:Weyl}

The displacement operator $D(z)$ can be expressed in terms of Weyl operators $W(q,p)$.
See Section 3.4.3 of \cite{GJP09}.
The latter have an easy definition by means of Stone's theorem.

The position operator $Q$ is the self-adjoint multiplication operator defined on
the Hilbert space ${\mathscr H}={\mathscr L}_2(\Ro,\Co)$ by $Q\psi(x)=x\psi(x)$.
The momentum operator $P$ is the self-adjont extension of the derivation operator
$-i\hbar\upd/\upd x$. The annihilation operator $a$ is the closed operator
defined by
\begin{equation}
a=\frac{1}{2r}\left(Q+i\frac{r^2}{\hbar}P\right).
\nonumber\end{equation}
Here, $r$ is an arbitrary positive constant with the dimension of a length.
The Weyl operators $W(q,p)$ are defined by
\begin{equation}
W(q,p)=e^{ipq/2\hbar}e^{ipQ/\hbar}e^{iqP/\hbar},
\qquad
q,p\in\Ro.
\nonumber\end{equation}
By definition is now
\begin{equation}
D(z)=W(-\sqrt 2 r z',\sqrt 2\hbar z''/r).
\nonumber\end{equation}

\section{Proof of Proposition \ref {disp:prop:deriv}}
\label{app:proofprop}

A short calculation gives
\begin{eqnarray}
D^\dagger(f)\partial_\mu\,D(f)
&=&
D^\dagger(f)\partial_\mu\,\left[
e^{i(f')(f'')}\,\,\,D(f')D(if'')\right]\cr
&=&
i\partial_\mu\,\left[(f')(f'')\right]\cr
& &
+D^\dagger(f)\exp\left(i(f')(f'')\right)\,\,\,
\left[\partial_\mu\,D(f')\right]D(if'')\cr
& &
+D^\dagger(f)\exp\left(i(f')(f'')\right)\,\,\,D(f')
\partial_\mu\,D(if'')\cr
&=& B_\mu
\label{calc:result}
\end{eqnarray}
with
\begin{equation}
B_\mu=
i\left(f\wedge \partial_\mu f(x)\,\right)
+\partial_\mu f(x)\,a^\dagger
-\partial_\mu\overline{f(x)}\,a.
\nonumber\end{equation}
Use $D(f)aD^\dagger(f)=a-f$ to obtain (\ref {calc:result2}).

Take now $\zeta$ in the domain of the number operator $a^\dagger a$.
Fix $\epsilon>0$ and introduce the abbreviations $z=f( x)$ and $w=f( x+\epsilon g_\mu)$.
For convenience, choose units in which $\hbar=1$ and $r=1$.
For simplicity take $\epsilon>0$.
One has
\begin{equation}
D(z)^\dagger D(w)\zeta
=
e^{-iw'w''}e^{-iz'z''}e^{ip_\epsilon Q}e^{iq_\epsilon P}\zeta
\label{app:proof:temp}
\end{equation}
with
\begin{equation}
q_\epsilon=-\sqrt 2(w'-z')
\quad\mbox{ and }\quad
p_\epsilon=\sqrt 2(w''-z'').
\nonumber\end{equation}
Let
\begin{eqnarray}
q_{,\mu}&=&-\sqrt 2 \partial_\mu z'\cr
p_{,\mu}&=&\sqrt 2 \partial_\mu z''.
\nonumber\end{eqnarray}
Rewrite (\ref {app:proof:temp}) in the following way
\begin{eqnarray}
D(z)^\dagger D(w)\zeta
&=&
e^{-iw'w''}e^{-iz'z''}e^{ip_\epsilon Q}
\left[e^{iq_\epsilon P}-1-i\epsilon q_{,\mu}P\right]\zeta\cr
&+&
e^{-iw'w''}e^{-iz'z''}
\left[e^{ip_\epsilon Q}-1-i\epsilon p_{,\mu}Q\right]
\left[1+i\epsilon q_{,\mu}P\right]\zeta\cr
&+&
\left[e^{-iw'w''}e^{-iz'z''}-1-i\epsilon (q_{,\mu}-p_{,\mu})/\sqrt 2 \right]\cr
& &\times
\left[1+i\epsilon p_{,\mu}Q\right]
\left[1+i\epsilon q_{,\mu}P\right]\zeta\cr
&+&
\left[1+i\epsilon (q_{,\mu}-p_{,\mu})/\sqrt 2 \right]
\left[1+i\epsilon p_{,\mu}Q\right]
\left[1+i\epsilon q_{,\mu}P\right]\zeta.
\nonumber\end{eqnarray}
Hence one has
\begin{eqnarray}
& &
||D(z)^\dagger D(w)\zeta 
-\left(1
+i\epsilon\left[
q_{,\mu} +p_{,\mu}\right]/\sqrt 2
++i\epsilon \left[p_{,\mu}Q+q_{,\mu}P \right]\right)\zeta||\cr
&\le&
||C_1||+||C_2||+|\lambda_3|\,||C_3||
+\epsilon^2 |q_{,\mu}+p_{,\mu}-1|\,||C_4||
+\epsilon^3|q_{,\mu}+p_{,\mu}|\,|p_{,\mu}q_{,\mu}|\,||C_5||\cr
& &
\label{app:proof:temp2}
\end{eqnarray}
with
\begin{eqnarray}
C_1&=&\left[e^{iq_\epsilon P}-1-i\epsilon q_{,\mu}P\right]\zeta\cr
C_2&=&\left[e^{ip_\epsilon Q}-1-i\epsilon p_{,\mu}Q\right]
\left[1+i\epsilon q_{,\mu}P\right]\zeta\cr
C_3&=&\left[1+i\epsilon p_{,\mu}Q\right]
\left[1+i\epsilon q_{,\mu}P\right]\zeta\cr
C_4&=&(p_{,\mu}Q+q_{,\mu}P)\zeta\cr
C_5&=&QP\zeta
\nonumber\end{eqnarray}
and
\begin{equation}
\lambda_3
=e^{-iw'w''}e^{iz'z''}-1-i\epsilon (q_{,\mu}+p_{,\mu})/\sqrt 2.
\nonumber\end{equation}

Let us now show that all contributions to the r.h.s.~of (\ref {app:proof:temp2})
are of order at most $\epsilon^2$.
In particular, $||C_1||$, $||C_2||$, $||C_3||$ and $\lambda_3$ must be estimated.
The estimates
are based on the inequalities
\begin{equation}
|e^{iu}-1|^2\le 2u^2
\quad\mbox{ and }\quad
|e^{iu}-1-iu|\le\frac 12u^2,
\qquad u\in\Ro.
\nonumber\end{equation}

\paragraph{$\bf ||C_1||$}
One has
\begin{eqnarray}
||C_1||
&=&
||(e^{iq_\epsilon P}-1-i\epsilon q_{,\mu} P)\zeta||\cr
&\le&
\frac 12 q_\epsilon^2|| P^2\zeta|| +|q_\epsilon-\epsilon q_{,\mu}|\,||P\zeta||.
\nonumber\end{eqnarray}
Note that $q_\epsilon=\mbox{O}(\epsilon)$
and $q_\epsilon-\epsilon q_{,\mu}=\mbox{o}(\epsilon)$.
By assumption is $\zeta$ in the domain of $P^2$ and hence of $P$.
One concludes that $||C_1||$ is of order strictly less than 1 in $\epsilon$.

\paragraph{$\bf ||C_2||$}
Similarly is
\begin{eqnarray}
||C_2||
&=&
||\left[e^{ip_\epsilon Q}-1-i\epsilon p_{,\mu}Q\right]
\left[1+i\epsilon q_{,\mu}P\right]\zeta||\cr
&\le&
||\left[e^{ip_\epsilon Q}-1-i\epsilon p_{,\mu}Q\right]\zeta||
+\epsilon|q,\mu|\,||\left[e^{ip_\epsilon Q}-1-i\epsilon p_{,\mu}Q\right]P\zeta||\cr
&\le&
\frac 12 p_\epsilon^2||Q^2\zeta||
+|p_\epsilon-\epsilon p_{,\mu}|\,||Q\zeta||
+\epsilon|q,\mu|\,||\left[e^{ip_\epsilon Q}-1\right]P\zeta||
+\epsilon^2|q_{,\mu}|\,|p_\epsilon|\,||QP\zeta||.
\nonumber\end{eqnarray}
Each of these terms is of order strictly less than 1 in $\epsilon$.

\paragraph{$||\bf C_3||$ and $\bf |\lambda_3|$}
One has
\begin{eqnarray}
||C_3||
&=&
||\left[1+i\epsilon p_{,\mu}Q\right]
\left[1+i\epsilon q_{,\mu}P\right]\zeta||\cr
&\le&
1+\epsilon |p_{,\mu}|\, ||Q\zeta||+\epsilon |q_{,\mu}|\, ||P\zeta||
+\epsilon^2 |p_{,\mu}q_{,\mu}|\,||QP\zeta||
\nonumber\end{eqnarray}
and
\begin{eqnarray}
|\lambda_3|
&=&
|e^{-iw'w''}e^{iz'z''}-1-i(\epsilon q_{,\mu}+p_{,\mu})/\sqrt 2 |\cr
&\le&
|e^{-iw'w''}e^{iz'z''}-1+i(w'w''-z'z'')|
+|w'w''-z'z''+\epsilon (q_{,\mu}-p_{,\mu})/\sqrt 2 |\cr
&\le&
\frac 12(w'w''-z'z'')^2+|w'w''-z'z''+\epsilon (q_{,\mu}-p_{,\mu})/\sqrt 2 |\cr
&=&\mbox{o}(\epsilon).
\nonumber\end{eqnarray}

\paragraph{Domain issues}

It follows from the Theorem of Rellich-Kato \cite{KT66}
that the operators $P^2$ and $a^\dagger a=(P^2+Q^2)/4-1/2$
have the same domain of definition.
By assumption does $\zeta$
belong to the domain of $a^\dagger a$ and hence of $P^2$ and of $Q^2$.
Finally, from $QP\zeta=i(P^2-Q^2-2)\zeta/2$ it follows that
$\zeta$ also belongs to the domain of the operator $QP$.

\section{Proof of Proposition \ref {diff:prop:coh}}
\label{app:coherent}

Here we show that if $\zeta$ is a coherent wave function then both $U(x)\zeta$
and $U(x)^\dagger\zeta$ are coherent wave functions times a complex phase factor. 

Let $\zeta=|z\ranglec$. One calculates
\begin{eqnarray}
U(x)\zeta
&=&
e^{i\phi(x)}D(f(x))^\dagger e^{i\gamma(x)a^\dagger a}D(f(x))|z\ranglec\cr
&=&
e^{i\phi(x)}e^{-i(f(x)\wedge z)}D(f(x))^\dagger e^{i\gamma(x)a^\dagger a}|z+f(x)\ranglec\cr
&=&
e^{i\phi(x)}e^{-i(f(x)\wedge z)}D(f(x))^\dagger|e^{i\gamma(x)}(z+f(x))\ranglec\cr
&=&
e^{i\phi(x)}e^{-i(f(x)\wedge z)}e^{i(f(x)\wedge(e^{i\gamma(x)}(z+f(x)))}
|e^{i\gamma(x)}(z+f(x))-f(x)\ranglec.\cr
& &
\label{app:coherent:expl}
\end{eqnarray}
This is a coherent wave function times a complex phase factor.

Similarly is
\begin{equation}
U^\dagger(x)\zeta
=
e^{-i\phi(x)}e^{-i(f(x)\wedge z)}e^{i(f(x)\wedge(e^{-i\gamma(x)}(z+f(x)))}
|e^{-i\gamma(x)}(z+f(x))-f(x)\ranglec.
\nonumber\end{equation}

\section{Proof of Proposition \ref {gauge:prop:difffield}}
\label {app:proofprop2}

For convenience, choose units in which 
$r=1$.
One has
\begin{eqnarray}
i\hbar \partial_\mu\,\eta(x)
&=&
i\hbar \partial_\mu\,
e^{i\phi(x)}
D(f(x))^\dagger e^{i\gamma(x)a^\dagger a}D(f(x))\zeta\cr
&=&
-\hbar(\partial_\mu \phi(x))\,\eta(x)\cr 
&-&
e^{i\phi(x)}D(f(x))^\dagger X_\mu(x)e^{i\gamma(x)a^\dagger a}D(f(x))\zeta\cr
&-&
\hbar (\partial_\mu\gamma(x))\, e^{i\phi(x)}
D(f(x))^\dagger a^\dagger a e^{i\gamma(x)a^\dagger a}D(f(x)) \zeta\cr
&+&
e^{i\phi(x)}D(f(x))^\dagger e^{i\gamma(x)a^\dagger a}X_\mu(x) D(f(x))\zeta
\nonumber\end{eqnarray}
with $X_\mu$ given by (\ref {disp:Xdef}).
Use
\begin{equation}
D(f(x))^\dagger X_\mu(x)D(f(x))
=
X_\mu+i\hbar \left(\partial_\mu f(x)\right)\overline{f(x)}
-i\hbar \left(\partial_\mu \overline{f(x)}\right)f(x) 
\nonumber\end{equation}
and
\begin{equation}
D(f(x))^\dagger a^\dagger aD(f(x))
=
(a+f(x))^\dagger(a+f(x))
\nonumber\end{equation}
and
\begin{eqnarray}
e^{i\gamma(x)a^\dagger a}X_\mu(x)e^{-i\gamma(x)a^\dagger a}
&=&
\hbar \left(f\wedge \partial_\mu f(x)\,\right)\cr
& &
+i\hbar e^{i\gamma(x)}\partial_\mu f(x)\,a^\dagger
-i\hbar e^{-i\gamma(x)}\partial_\mu \overline{f(x)}\,a.
\nonumber\end{eqnarray}
to obtain
\begin{equation}
i \partial_\mu\,\eta(x)
=
R_\mu(x)\eta(x)
\nonumber\end{equation}
with $R_\mu$ given by 
\begin{eqnarray}
R_\mu(x)
&=&
-\hbar\partial_\mu\phi(x) -2\hbar f(x)\wedge\partial_\mu f(x)
+i\hbar \partial_\mu f(x) \, a^\dagger
-i\hbar \partial_\mu\overline{f(x)}\, a\cr
& &
-\hbar(\partial_\mu\gamma(x))\,(a+f(x))^\dagger(a+f(x))\cr
& &
+i\hbar (\partial_\mu f(x)) e^{i\gamma(x)}\, (a+f(x))^\dagger
-i\hbar (\partial_\mu\overline{f(x)})e^{-i\gamma(x)}\, (a+f(x)).
\nonumber\end{eqnarray}
The latter expression reduces to (\ref {diff:Rpot}).
This proves part of item (b) of the proposition.

From the definition of the map $x\mapsto \eta(x)$ one sees that
it is a concatenation of Fr\'echet-differentiable maps.
The map $\zeta\mapsto D(f(x))\zeta$ is Fr\'echet-differentiable by
Proposition \ref {disp:prop:deriv}.
The Fr\'echet-differentiability of $x\mapsto e^{i\gamma(x)a^\dagger a}$
follows from the estimate
\begin{eqnarray}
& &
||\left[e^{i\gamma(y)a^\dagger a}
-e^{i\gamma(x)a^\dagger a}
+i\gamma'(x)a^\dagger ae^{i\gamma(x)a^\dagger a}\right]\zeta||\cr
&=&
||\left[1-e^{i[\gamma(x)-\gamma(y)]a^\dagger a}
+i\gamma'(x)a^\dagger a e^{i[\gamma(x)-\gamma(y)]a^\dagger a}
\right]\zeta||\cr
&\le&
||\left[1-e^{i[\gamma(x)-\gamma(y)]a^\dagger a}
+i[\gamma(x)-\gamma(y)]a^\dagger a \,e^{i[\gamma(x)-\gamma(y)]a^\dagger a}\right]\zeta||\cr
& &
+|\gamma'(x)-\gamma(x)+\gamma(y)|\,||a^\dagger a\,\zeta||\cr
&\le&
\frac 12|\gamma(x)-\gamma(y)|^2\,||a^\dagger a\,\zeta||^2\cr
& &
+|\gamma'(x)-\gamma(x)+\gamma(y)|\,||a^\dagger a\,\zeta||\cr
&=&\mbox{ o}(|x-y|).
\nonumber\end{eqnarray}
This shows item (a) of the proposition.

Let us finally finish the proof of item (b) of the proposition.
The r.h.s.~of (\ref {diff:Rpot}) can be written as
\begin{equation}
R_\mu=\lambda (P+\mu)^2 +\lambda (Q+\nu)^2+\rho
\nonumber\end{equation}
with 
\begin{equation}
\lambda=-\frac 14\partial_\mu\gamma(x)\,
\nonumber\end{equation}
and further real constants $\mu,\nu,\rho$.
By the theorem of Rellich-Kato \cite{KT66} the operators $R_\mu$
and $(P+\mu)^2$ have the same domain of definition and $R_\mu$ is a self-adjoint operator.
The domain of definition of $P^2$ is $\mathscr D$. Hence, this is also the
domain of definition of $R_\mu$.

\section{Orthogonal fields}
\label{app:ortho}

From (\ref {diff:antisym}) one obtains 
\begin{eqnarray}                                    
\Ee
&=&
-i\hbar\left(e^{i\gamma(x)}-1\right)\left(e^{-i\gamma(x)}-1\right)\,\left[
\left(\partial_0 f(x)\right)\,\overline {\nabla f}
-\left(\partial_0 \overline{f(x)}\right)
\,\nabla f
\right]\cr
& &
-\hbar\left(e^{-i\gamma(x)}-1\right)\,
\left[
\left(\partial_0\gamma(x)\right)\,\, \nabla f
-\left(\partial_0 f(x)\right)\, \nabla \gamma
\right]\,
(a+f(x))^\dagger
\cr
& &
-\hbar\left(e^{i\gamma(x)}-1\right)\,
\left[
\left(\partial_0\gamma(x)\right)\,\,\overline {\nabla f}
-\left(\partial_0 \overline{f(x)}\right)\,\nabla\gamma
\right]\,
(a+f(x))
\nonumber\end{eqnarray}
and
\begin{eqnarray}
\Bb
&=&
-i\hbar\left(e^{i\gamma(x)}-1\right)\left(e^{-i\gamma(x)}-1\right)\,
\overline{\nabla f}\times\nabla f\cr 
& &
-\hbar \left(e^{-i\gamma(x)}-1\right)\,
\left(\nabla f\times \nabla\gamma\right)\,(a+f(x))^\dagger\cr
& &
-\hbar \left(e^{i\gamma(x)}-1\right)\,
\left(\overline{\nabla f}\times\nabla\gamma\right)\,(a+f(x)).
\nonumber\end{eqnarray}

The conditions for $\Ee$ and $\Bb$ to be orthogonal to $\kk$ become
\begin{eqnarray}
0
&=&
\left(\partial_0 f(x)\right)\,\kk\cdot\overline{\nabla f}
-
\left(\partial_0 \overline{f(x)}\right)\,\kk\cdot\nabla f,\cr
0
&=&\left(\partial_0 f(x)\right)\,
\kk\cdot\nabla\gamma
-
\left(\partial_0\gamma(x)\right)\,\,
\kk\cdot\nabla f,\cr
0
&=&
\kk\cdot\,(\nabla f\times\overline{\nabla f}),\cr           
0
&=&
\kk\cdot\,(\nabla f\times\nabla\gamma).
\nonumber\end{eqnarray}
If $\gamma$ and $f$ satisfy these equations then there exist complex numbers $z$ and $w$
such that
\begin{equation}
\nabla f
=
z\nabla\gamma+w\kk.
\nonumber\end{equation}
The expression for the magnetic field simplifies to
\begin{eqnarray}
\Bb
&=&
-i\hbar\left(e^{i\gamma(x)}-1\right)\left(e^{-i\gamma(x)}-1\right)\,
(z\overline w-w\overline z)\,\left(\kk\times\nabla\gamma\right)\cr 
& &
-\hbar \left(e^{-i\gamma(x)}-1\right)\,
w\,\left(\kk\times\nabla\gamma\right)\,(a+f(x))^\dagger\cr
& &
-\hbar \left(e^{i\gamma(x)}-1\right)\,
\overline w\, \left(\kk\times\nabla\gamma\right)\,(a+f(x)).
\nonumber\end{eqnarray}
The orthogonality conditions simplify to
\begin{eqnarray}
0
&=&
\left(\partial_0 f(x)\right)
\left(\overline z\,\kk\cdot {\nabla \gamma}+\overline w |\kk|^2\right)
-
\left(\partial_0 \overline{f(x)}\right)\left(z\,\kk\cdot\nabla \gamma+w|\kk|^2\right),\cr
0
&=&\left(\partial_0 f(x)\right)\,
\kk\cdot\nabla\gamma
-
\left(\partial_0\gamma(x)\right)\,
\left(z\,\kk\cdot\nabla \gamma+w|\kk|^2\right).
\nonumber\end{eqnarray}

First assume that $\kk\cdot\nabla\gamma=0$.
Then the remaining requirements are
\begin{equation}
\overline w \left(\partial_0 f(x)\right)\mbox{ is real and }
w\,\partial_0\gamma(x)\,=0.
\nonumber\end{equation}
In this case the expression for the electric field becomes
\begin{eqnarray}
\Ee
&=&
\nabla\gamma\cr
&\times&\bigg[
i\hbar\left(e^{i\gamma(x)}-1\right)\left(e^{-i\gamma(x)}-1\right)\,\left[
z \partial_0 \overline{f(x)}-\overline z \partial_0 f(x)\right]\cr
& &
-\hbar\left(e^{-i\gamma(x)}-1\right)\,
\left[
z \partial_0\gamma(x)-\partial_0 f(x)\right]\,
(a+f(x))^\dagger
\cr
& &
-\hbar\left(e^{i\gamma(x)}-1\right)\,
\left[
\overline z \partial_0\gamma(x)-\partial_0 \overline{f(x)}\,\right]\,
(a+f(x))
\bigg].
\nonumber\end{eqnarray}

On the other hand, if $\kk\cdot\nabla\gamma\not=0$ then the only remaining requirement is
condition (\ref {app:ortho:timecond}).
The expressions for the electric field becomes
\begin{eqnarray}
\Ee
&=&
\left(\frac{|\kk|^2}{\kk\cdot\nabla\gamma}\nabla\gamma-\kk\right)\cr
&\times&
\bigg[
i\hbar\left(e^{i\gamma(x)}-1\right)\left(e^{-i\gamma(x)}-1\right)\,
\left(\partial_0\gamma(x)\right)\,(z\overline w-w\overline z)
\cr
& &
+\hbar\left(e^{-i\gamma(x)}-1\right)\,
w \left(\partial_0\gamma(x)\right)\,
(a+f(x))^\dagger\cr
& &
+\hbar\left(e^{i\gamma(x)}-1\right)\,
\overline w \left(\partial_0\gamma(x)\right)\,
(a+f(x))
\bigg].
\nonumber\end{eqnarray}

Select for instance any vector $\hh$ orthogonal to $\kk$
and let $h_0=|\kk|$ and $\gamma(x)=(k_\mu +h_\mu)x^\mu$.
Take $z=0$ as it appears only in the scalar contributions to the fields.
Then one finds
\begin{equation}
\nabla\gamma=\kk+\hh
\quad\mbox{ and }\quad
\partial_0\gamma(x)\,=2|\kk|.
\nonumber\end{equation}
The function $f$ should satisfy
\begin{equation}
\nabla f=w\kk
\quad\mbox{ and }\quad
\partial_0 f(x)=2|\kk|w,
\nonumber\end{equation}
with $w$ a complex number.
The expressions for the field vectors become
\begin{eqnarray}
\Ee&=&2\hbar|\kk|\,\hh\,\left(e^{-i(k_\mu+h_\mu)x^\mu}-1\right)w(a+f(x))^\dagger
+\mbox{h.c.},\cr
\Bb&=&-\hbar \,\left(\kk\times\hh\right)\,\left(e^{-i(k_\mu+h_\mu)x^\mu}-1\right)w(a+f(x))^\dagger
+\mbox{h.c.}.
\nonumber\end{eqnarray}

\newpage
\section*{}

\end{document}